\documentclass[prb,twocolumn,amsmath,aps,showpacs,superscriptaddress]{revtex4}
\usepackage{graphicx}
\usepackage{dcolumn}
\usepackage{bm}
\usepackage{hyperref}

\begin{document}

\title{Width of the $0-\pi$ phase transition in diffusive magnetic Josephson junctions}

\author{Zahra Shomali}

\author{Malek Zareyan}

\affiliation{Institute for Advanced Studies in Basic Sciences
(IASBS), P.O. Box 45195-1159, Zanjan 45195, Iran }

\author{Wolfgang Belzig}

\affiliation{
Fachbereich Physik, Universit¨at Konstanz, D-78457 Konstanz, Germany
}

\date{\today}
%
\begin{abstract}
  We investigate the Josephson current between two superconductors (S)
  which are connected through a diffusive magnetic junction with a
  complex structure (F$_{c}$). Using the quantum circuit theory, we
  obtain the phase diagram of $0$ and $\pi$ Josephson couplings for
  F$_{c}$ being a IFI (insulator-ferromagnet-insulator) double barrier
  junction or a IFNFI structure (where N indicates a normal metal
  layer). Compared to a simple SFS structure, we find that the width
  of the transition, defined by the interval of exchange fields in
  which a $0-\pi$ transition is possible, is increased by insulating
  barriers at the interfaces and also by the presence of the
  additional N layer. The widest transition is found for symmetric
  F$_{c}$ structures. The symmetric SIFNFIS presents the most
  favorable condition to detect the temperature induced $0-\pi$
  transition with a relative width, which is five times larger than
  that of the corresponding simple SFS structure.
\end{abstract}
%

\pacs{74.45.+c, 74.50.+r, 72.25.-b}

\maketitle

%
\section{\label{sec:intro}Introduction}
Ferromagnet-superconductor (FS) heterostructures feature novel and
interesting phenomena, which have been active topics of investigation
for more than half a century \cite{GKI04,BVE05,Buz05}.  Meanwhile,
Josephson structures comprising a ferromagnetic weak link have been
studied extensively. The existence of the $\pi$-junction in a SFS
  structure is one of the most interesting phenomena which occurs for
  certain thicknesses and exchange fields of the F layer
  \cite{BKS77,Buz82,ROR01,ROVR01,KALG02,GABKL03,SBLC03,BPSB08,CBNB01,BVE01,
    RLF03,Cht04,BK91,Buz03,MZ06,VGKW08,LYS08}. This
  manifestation is due to the oscillatory behavior of the
  superconducting pair amplitude and the electronic density of states
  in the ferromagnet \cite{Buz00,ZB01,ZB02}. In a $\pi$-junction the
ground state phase difference between two coupled superconductors is
$\pi$ instead of $0$ as in the usual $0$-state SNS junctions. The
existence of a $\pi$-state was predicted theoretically by Bulaevski
$\emph{et al.}$ \cite{BKS77} and Buzdin $\emph{et al.}$ \cite{Buz82},
and has been first observed by Ryazanov $\emph{et al.}$
\cite{ROR01}. The transition from the $0$- to $\pi$-state is
associated with a sign change of the critical current, $I_{c}$, which
leads to a cusp-like dependence of the absolute values of $I_{c}$ on
temperature. Later, the nonmonotonic temperature dependence of the
critical current in diffusive contacts was observed in other
experiments \cite{ROVR01,KALG02,GABKL03,SBLC03,BPSB08} and was
attributed to the $0-\pi$ transition induced by the ferromagnetic
exchange field. The $0-\pi$ transition has been studied theoretically
by several authors in clean \cite{CBNB01,BVE01,RLF03,Cht04} and
diffusive \cite{BK91,BVE01,Buz03,Cht04,MZ06,VGKW08,LYS08} Josephson
contacts for different conditions and barriers at the FS interfaces.

An interesting application of a $\pi$-junction is a superconducting
qubit as one of the most noticeable candidates for a basic element
of quantum computing. Furthermore, $\pi$-junctions have been
proposed as phase qubit elements in superconducting logic circuits
\cite{IGF99,MOL99,BRG03,YTTM05,YTM06}. Also, a phase qubit in SIFIS
junctions, in which the qubit state is characterized by the $0$ and the
$\pi$ phase states of the junction, has recently been suggested
\cite{NKS08}. Due to these exciting proposed applications, the
detection of $0-\pi$ transitions with very high sensitivity is
necessary. Investigating the phase diagrams of $0-\pi$ transitions
\cite{CBNB01,LCE07} for different structures with different
characteristics should make it possible to determine the most efficient
control of the $0-\pi$ transition.

In this paper we investigate the width of the temperature-induced
$0-\pi$ transition in a diffusive SF$_{c}$S junction. Here, F$_{c}$
represents a complex ferromagnetic junction of length $L$, which
consists of diffusive ferromagnetic and normal metallic parts as
well as insulating barriers. We define the width of the transition
as the interval of exchange fields, in which the
temperature-dependent transition from the $0$- to $\pi$-phase is
possible. We use the so-called quantum circuit theory (QCT), which
is a finite-element technique for quasiclassical Green's functions
in the diffusive limit \cite{Naz94,Naz99,Naz05}. The QCT-description
has the advantage, that it does not require to specify a concrete
geometry. By a discretization of the Usadel equation \cite{Usa70}
one obtains relations analogous to the Kirchhoff laws for classical
electric circuit theory. These relations can be solved numerically
by iterative methods and one obtains the quasiclassical Green's
function of the whole system. The QCT has been generalized to
spin-dependent transport in Ref.~\onlinecite{HHNB02p,HHNB02}.
We adopt the finding of this paper for FN contacts to handle our
problem of the SF$_{c}$S contacts. We discretize the interlayer
between the superconducting reservoirs into nodes. Following
Refs.~\onlinecite{HHNB02,BN07}, every node in a ferromagnetic layer
with specific exchange field is equivalent to a normal node
connected to a ferromagnetic insulator reservoir which determines
the exchange field. This similarity has been verified experimentally
with EuO$|$Al$|$Al$_{2}$O$_{3}|$Al junctions \cite{TTK86}. It has
been found that the induced exchange-field of the EuO-insulator,
which is responsible for spin-splitting in the measured density of
states, was of the same order as its magnetization \cite{TTK86}.
Also, the authors of Ref.~\onlinecite{HHNB02} have shown that the
normalized density of states in the normal metal, which is connected
to a superconductor and an insulator ferromagnet at its ends, is the
same as the one for a BCS superconductor in the presence of a
spin-splitting magnetic field \cite{GPTFS80,MT94}. This method
allows us to calculate the Josephson current flowing through the
SF$_{c}$S contact for an arbitrary length $L$ and all temperatures
while fully taking into account the nonlinear effects of the induced
superconducting correlations.

We investigate the width of the transition, $\Delta h$, for four
different cases of SF$_{c}$S structures with ideally transparent
FS-interfaces, symmetric SIFIS and asymmetric SIFS double barrier
F-junctions, and more complicated SIFNFIS structures (where I and N
denote, respectively, insulating barrier and normal metal). For a
fixed length $L$, all these systems show several transition lines in
the phase diagram of $T/T_{c}$ and $h/T_{c}$. Higher order
transitions occur at large exchange fields $h$. We find that higher
order transitions are wider than the first transition. Also,
decreasing the contact length $L$ leads to a widening of the
transitions and, at the same time, to an increasing of the exchange
field, $h_{in}$, at which the transition starts. Nevertheless, the
relative width of the transition, given by the ratio $\Delta
h/h_{in}$, decreases.

For the SIFIS structure we show that the existence of the I-barriers
at the FS interfaces broadens the $0-\pi$ transitions and, hence,
improves the conditions to detect such transitions. In addition, we
find that a symmetric double barrier structure with the two barriers
having the same conductance shows wider transitions than the
corresponding asymmetric structure with the same total conductance
but different conductances of the barriers. An even larger width of
transitions can be achieved by including an additional normal metal
part into F$_{c}$. This motivates our study of an SIFNFIS structure,
for which relative width $\Delta h/h_{in}$ is in general larger than
that of the corresponding SIFIS with the same total conductance and
the mean exchange field of the F$_{c}$ part.

The structure of this paper is as follows. In Sec. \ref{sec:level1},
we introduce the model and basic equations, which are used to
investigate the SF$_{c}$S-Josephson junction. We introduce the
finite-element description of our structures using quantum circuit
theory technique. In Sec.~\ref{sec:level2}, we investigate phase
diagrams of $0-\pi$ transitions for the SFS, SIFIS, and SIFNFIS
structures. Analyzing our findings, we determine the most favorable
conditions for an experimental detection of the $0-\pi$ transitions.
Finally, we conclude in Sec.~\ref{sec:level3}.
%
\section{\label{sec:level1}Model and Basic Equations}

\begin{figure}
\begin{center}
\includegraphics[width=3.1in]{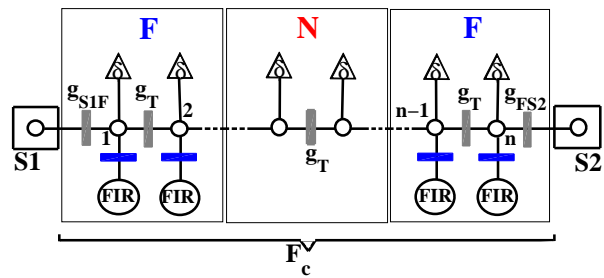}
\end{center}
\caption{\label{Fig:1}(Color online) Quantum circuit model of the
SF$_{c}$S structure. The contact $F_{c}$ is discretized into $n$
nodes which are connected to each other by tunnel barriers of
conductance $g_{T}$; $g_{S1F}$ and $g_{FS2}$ denote the
conductances of S1F and FS2 interfaces, respectively.  The inverse
of the average level spacing, $\delta$, represents the leakage
term due to a finite volume of a node; FIR represents a
ferromagnetic insulating reservoir.}
\end{figure}

We consider a ferromagnetic SF$_{c}$S Josephson structure in which
two conventional superconducting reservoirs are connected by a
complex diffusive F$_{c}$ junction. We investigate three cases of
F$_{c}$: (i) a simple F layer with a homogeneous spin-splitting
exchange field $h$ (SFS), (ii) a double barrier IFI structure, in
which the F-layer is connected via I-barriers to the reservoirs
(SIFIS), and (iii) an IFNFI junction composed of two ferromagnetic
layers with the same length $L_{F}$ and the same exchange field and
a normal metal with length $L_{N}$ in between such that
$L=L_{N}+2L_{F}$ (SIFNFIS). We compare the width of the
temperature-induced $0-\pi$ transitions for these three types of
structures. In all cases F$_{c}$ has the same length, total
conductance, and the mean exchange field $h$.

In our approach, we make use of the quantum circuit theory which is
a finite-element theory technique for quasiclassical Green's
function method in diffusive limit \cite{Naz94,Naz99,Naz05}. In this
technique, each part of the structure is represented by a node which
is connected to other nodes or superconductor/ferromagnet reservoirs
\cite{HHNB02}. Green's functions are calculated by using balance
equations for matrix currents entering from the connectors, which is
described in terms of its transmission properties and Green's
functions of the nodes forming it, to each node. For calculations we
follow the procedure similar to that of Ref.~\onlinecite{BN07}. We
discretize the conducting part of $F_{c}$ into $n$ nodes as
presented in Fig.~\ref{Fig:1}. A node in the ferromagnetic part will
be presented by a normal-metal node connected to a ferromagnetic
insulating reservoir (FIR) which induces an exchange field equal to
the exchange field of the ferromagnetic part at the place of the
node.

Each of the superconducting reservoirs is assumed to be a standard
BCS-superconductor. Our circuit connecting those reservoirs consists
of different types of nodes in F$_{c}$. One type are the normal nodes
in the middle of F$_{c}$, each of which is connected to two
neighboring nodes which are either normal nodes or F nodes. Another
type are F nodes placed at the two ends of F$_{c}$, where each of
them, in addition to its connection to two neighboring nodes, is
connected to FIR as well and, hence, feels the exchange field
directly. As can be seen in Fig.~\ref{Fig:1}, each of the two
neighboring nodes of a F node can be another F node, an N node, or a
superconducting node. We denote the conductances of the tunnel
  barriers at S1F and FS2 interfaces by $g_{S1F}$ and $g_{FS2}$,
  respectively. Also, $g_{T}$ represents the conductance of the tunnel
  barrier between each two nodes inside $F_{c}$; $g_{T}$ is determined
  by $g_{F_{c}}$, the total conductance of $F_{c}$, excluding the
  conductances of the barriers at the interfaces
  $(n-1)/g_{T}=(1/g_{F_{c}})-(1/g_{S1F}+1/g_{FS2})$. In general, a node
$i$ is characterized by a Green's function $\check{G}_{i}$, which is
an energy-dependent $4\times4$-matrix in the Nambu and spin spaces.
Furthermore, all nodes in F$_{c}$ are assumed to be coupled to each
other by means of tunneling contacts. However, a finite volume of a
node and the associated decoherence between electron and hole
excitations are taken into account by the leakage matrix current which
is proportional to the energy, $\epsilon$, and the inverse of the
average level spacing in the node, $\delta$ \cite{Naz94,Naz99}.

For a structure with spin-dependent magnetic contacts and in the
presence of F and S reservoirs, the matrix current was developed in
Ref. \onlinecite{HHNB02}. In the limit of tunneling contacts, which is
our interest, the matrix current between two nodes $i, j$ is defined
as \cite{HHNB02,HN05},
\begin{equation}
\label{eq:mxcurrent}
\check{I}_{i,j}=\frac{g_{i,j}}{2}[\check{G}_{i},\check{G}_{j}]+\frac{G_{MR}}{4}[\{
(\vec{h}_{i}.\vec{\hat{\sigma}})\hat{\tau}_{3},\check{G}_{i}\},\check{G}_{j}]
\end{equation}
\begin{equation}
\nonumber\\
 +[i\frac{G_{Q}}{\delta_{i}}(\vec{h}_{i}.\vec{\hat{\sigma}})\hat{\tau}_{3},\check{G}_{j}].
\end{equation}

The first term demonstrates the usual boundary condition for a
tunneling junction, where $g_{i,j}$ is the tunneling conductance of
the contact between the two nodes. The second term exists due to the
different conductances for different spin directions which leads to
the spin polarized current through the contact. We assume this term
to be negligible as, $G_{MR}\sim
g^{\uparrow}_{i,j}-g^{\downarrow}_{i,j} \ll g_{i,j}$. Also,
$G_{Q}\equiv e^{2}/2\pi\hbar$ is the quantum of conductance,
$\vec{h}$ is the exchange field of the node, and $\vec{\sigma}$ and
$\vec{\tau}$ are the vectors consisting of Pauli matrices in spin
and Nambu space.

Using Eq. (\ref{eq:mxcurrent}) for different matrix currents
entering into a given node $i$, we apply the condition of current
conservation to obtain the following balance equation,
\begin{equation}
\label{eq:balance}
[\sum_{j=i-1,i+1}\frac{g_{j,i}}{2}\check{G}_{j}+i\frac{G_{Q}}{\delta_{i}}
(\vec{h}_{i}.\vec{\hat{\sigma}})\hat{\tau}_{3}-i\frac{G_{Q}}{\delta_{i}}\epsilon\hat{\tau}_{3}\hat{\sigma}_{0},\check{G}_{i}]=0.
\end{equation}
Here, the first term represents the matrix currents from neighboring
nodes $i-1, i+1$, which could be F, N or S. The second and third
terms are, respectively, the exchange term and the leakage matrix
current. Also, $\hat{\sigma_{0}}$ represents unit matrix in spin
space.

We consider the case, in which the exchange field in the ferromagnetic
parts of F$_{c}$ is homogeneous and collinear. Then, it is sufficient
to work with the $2\times2$ matrix Green's function of spin-$\sigma$
($\sigma=\uparrow/\downarrow$) electrons in Nambu space.  In the
Matsubara formalism the energy $\varepsilon$ is replaced by Matsubara
frequency $i\omega=i\pi T(2m+1)$) and the Green's function has the
form
\begin{equation}
\hat{G}=\left(
    \begin{array}{cc}
      G & F \\
      F^{*} & -G
    \end{array}
  \right).
\end{equation}

Neglecting the inverse proximity effect in the right and left
superconducting reservoirs, we set the boundary conditions at the
corresponding nodes S1 and S2 to the bulk values of the matrix Green's
functions:
\begin{equation}
\hat{G}_{1,2}=\frac{1}{\sqrt{\omega^{2}+\Delta^{2}}}\left(
  \begin{array}{cc}
    \omega & \Delta e^{\pm i\phi/2} \\
    \Delta e^{\mp i\phi/2} & -\omega \\
  \end{array}
\right)\,.
\end{equation}
Here $\Delta e^{\pm i\phi/2}$ are, respectively, the superconducting
order parameters in the right and left superconductors, and $\phi$
is the phase difference. The matrix Green's function satisfies
the normalization condition, $\hat{G}^{2}=\hat{1}$. The
temperature-dependence of the superconducting gap $\Delta$ is
modeled by the following formula \cite{Muh59,Bel99}
\begin{equation}
\Delta=1.76T_{c}\textrm{tanh}(1.74\sqrt{\frac{T}{T_{c}}-1}).
\end{equation}

We scale the size of F$_{c}$ in units of the diffusive
superconducting coherence length, $\xi_{S}=\sqrt{\xi_{0}l_{imp}}$
where $\xi_{0}=v_{F}/\pi \Delta_{0}$ with $v_{F}$ being the Fermi
velocity and $\Delta_{0}=\Delta(T=0)=1.76T_{c}$, and $l_{imp}$ is
the mean free path in the F-layer related to the diffusion
coefficient via $D=v^{(F)}_{F}l_{imp}/3$. Two more scales that we
use are $h/T_{c}$ and $T/T_{c}$, where $T_{c}$ is the critical
temperature of S reservoirs. Also, the mean level spacing depends on
the size of the system via the Thouless energy
$E_{Th}=D/L^{2}\equiv g_{T}\delta/(n-1)G_{Q}$ (Planck and Boltzmann
constants, $\hbar$ and $k_{B}$, are set to 1 throughout this paper).

In the absence of spin-flip scatterings, the balance equation, Eq.
(\ref{eq:balance}), is written for each spin direction separately for
all $n$ nodes in F$_{c}$. This results in a set of equations for $n$
matrix Green's functions of the nodes that are solved numerically by
iteration. In our calculation we start with choosing a trial form of
the matrix Green's functions of the nodes, for a given $\phi$, $T$,
and the Matsubara frequency $m=1$. Then, using
  Eq. (\ref{eq:balance}) and the boundary conditions iteratively, we
  refine the initial values until the Green's functions are calculated
  in each of $n$ nodes with the desired accuracy. Note that in general
  for any phase differennce $\phi$, the resulting Green's functions
  vary from one node to another, simulating the spatial variation
  along the F$_c$ contact.  From the resulting Green's functions we
calculate the spectral current using Eq.  (\ref{eq:mxcurrent}) and
obtain
\begin{equation}
\label{eq:3.12} I=\frac{T}{4e}(2\pi
i)\sum_{\omega_{m}=-\infty}^{\omega_{m}=\infty}Tr(\hat{\tau}_{3}\hat{I}).
\end{equation}

In the second step we set the next Matsubara frequency $m=2$, find
its contribution to the spectral current, and continue to the higher
frequencies until the required precision of the summation over $m$
is achieved. Finding the spectral current, for the given temperature
and phase difference, enables us to obtain the dependence of the
critical current $I_{c}$ on $T$. Finally, we increase the number of
nodes, $n$, and repeat the above procedure until all the spectral
currents for every temperature and phase difference reach the
specified accuracy. We find that for typical values of the involved
parameters, a mesh of 60 nodes is sufficient to obtain $I_{c}$
through the diffusive F$_{c}$ structure with an accuracy of
$10^{-3}$ across the whole temperature range.

%
\section{\label{sec:level2}Results and Discussions}

>From the numerical calculations, described above, we have obtained
the phase diagram of $0-\pi$ transition in the plane of $h/T_{c}$
and $T/T_{c}$. We analyze the width of $0-\pi$ transitions for the
SFS, symmetric SIFIS and asymmetric SIFS double barrier junctions,
and SIFNFIS structures.

Concerning such transitions, the width $\Delta h$ defines the
interval of the $h$, in which a temperature-induced transition
occurs. We compare relative width, the ratio $\Delta h/h_{in}$, of
different structures, where $h_{in}$ is the exchange field in which
the transition starts (see Fig.~\ref{Fig:2}a). In practice, we fix
the size of the structures, $L/\xi_{s}$, and then vary the value of
$h/T_{c}$ for detecting the change in the sign of the critical
supercurrent as the transition occurs. We expect that the detection
of a $0-\pi$ transition can be more feasible for the structure
having larger $\Delta h/h_{in}$.

\subsection{SFS structures}
\begin{figure}
\centering
\includegraphics[width=\columnwidth]{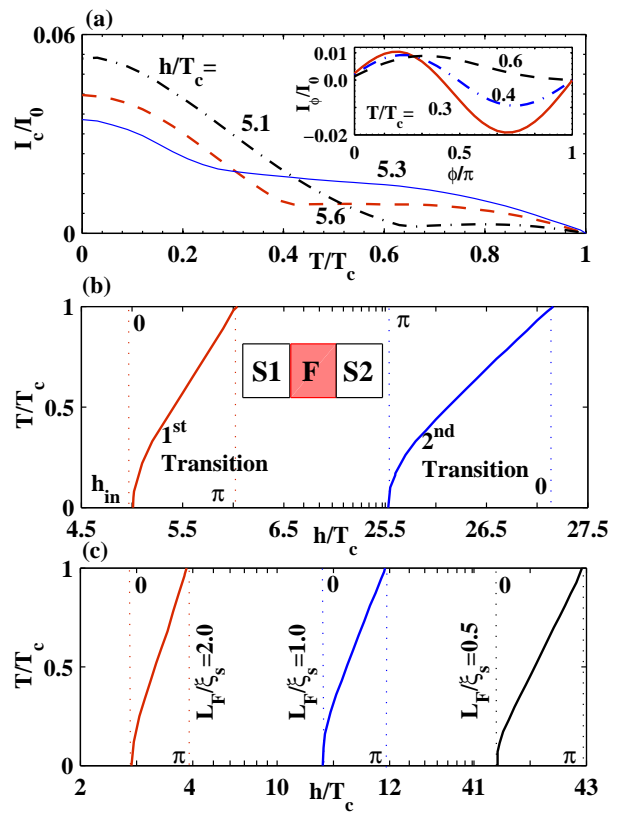}
\caption{\label{Fig:2}(a) (Color online) Normalized critical
current, $I_c/I_0$, versus temperature for a SFS structure with
the length $L_F/\xi_s=1.5$ and different exchange fields
$h/T_{c}$. The inset shows the current-phase characteristic for
$h/T_{c}=5.1$ in the vicinity of the $0-\pi$ transition
temperature. (b) The corresponding $0-\pi$ phase diagram showing
the phase boundaries up to the second transition. (c) Phase
diagram of the first $0-\pi$ transition for different length of
the junction $L_F/\xi_{s}$.}
\end{figure}

First, we consider the SFS structure. Figure.~\ref{Fig:2}a presents
the typical 0-$\pi$ transitions for such a junction with
$L/\xi_{s}=1.5$, where the supercurrent is scaled in units of
$I_{0}=(\pi/2)\Delta_{0}/eR_{F_{c}}$. Here, $R_{F_{c}}$ is the total
resistance of F$_{c}$. We observe that the nonzero supercurrent
  at the transition point is larger when the transition temperature is
  lower. Also, the phase diagram is shown in Fig.~\ref{Fig:2}b in the
vicinity of the first and the second $0-\pi$ transitions. At the first
transition the junction goes from the $0$- to the $\pi$-state starting at
$h_{in}$ and $T=0$. Increasing $h$, the transition temperature
increases toward $T_{c}$ and above the value $h=h_{in}+\Delta h$,
the junction will be in $\pi$ state at all temperatures. Increasing
$h$ further, the junction stays at its $\pi$ state until the exchange
field reaches the value at which the second transition starts (see
Fig.~\ref{Fig:2}b) where the junction changes back to a $0$-state. In
principal, it is possible to go to the higher exchange fields to see
higher transitions. However the amplitude of the supercurrent will be
extremely small and difficult to detect experimentally.

We have observed that the second transition is always wider than the
first one. In the case of Fig.~\ref{Fig:2}b, the width of the first
transition is nearly 0.65 of that of the second one. Furthermore,
the relative width for first transition is 0.20, while the second
transition has $\Delta h/h_{in}=0.06$. This finding can also be
generalized to higher transitions. In brief, higher transitions are
always associated with larger widths. In spite of having a smaller
width, the first transition seems to be more feasibly detectable, since
they have higher $\Delta h/h_{in}$.

Looking at the origin of the existence of $0-\pi$ transition, we can
understand this finding. An oscillating behavior of the order
parameter in a ferromagnetic layer makes the occurrence of different
signs of order parameters of the superconductor reservoirs,
possible. This effect, being in charge of the $\pi$-phase state, can
be seen when the length of the ferromagnet is of the order of
half-integer multiple of a period $2\pi \xi_{F}$, where $\xi_{F}$ is
the ferromagnetic coherence length of the ferromagnet. In the
diffusive limit when $h>>T_{c}$, this length is equal to
$\xi_{F}=\sqrt{D/h}$.  Hence, as $d\xi_{F}/dh$ is inversely
proportional to the exchange field, when the exchange field becomes
larger the rate of reduction of $\xi_{F}$ decreases and the system
will remain longer in the region of transition.

Now let us consider the effect of the length of F on the width of
the transition. In Fig.~\ref{Fig:2}c, the width of the first
transition for three lengths $L/\xi_{s}=0.5, 1, 2$ are compared. As
mentioned above, the condition for the occurrence of the first
transition is that the length $L$ becomes of the order of
half-integer of the period. For a smaller $L$ this condition is
fulfilled at larger $h$ which, in light of the above discussion,
means a wider $0-\pi$ transition. This can be seen easily in Fig.
~\ref{Fig:2}c. Note that the transition between the two states
always starts from lower temperatures.

\subsection{SIFIS, SIFS, and SIFNFIS structures}
\begin{figure}
\begin{center}
\includegraphics[width=\columnwidth]{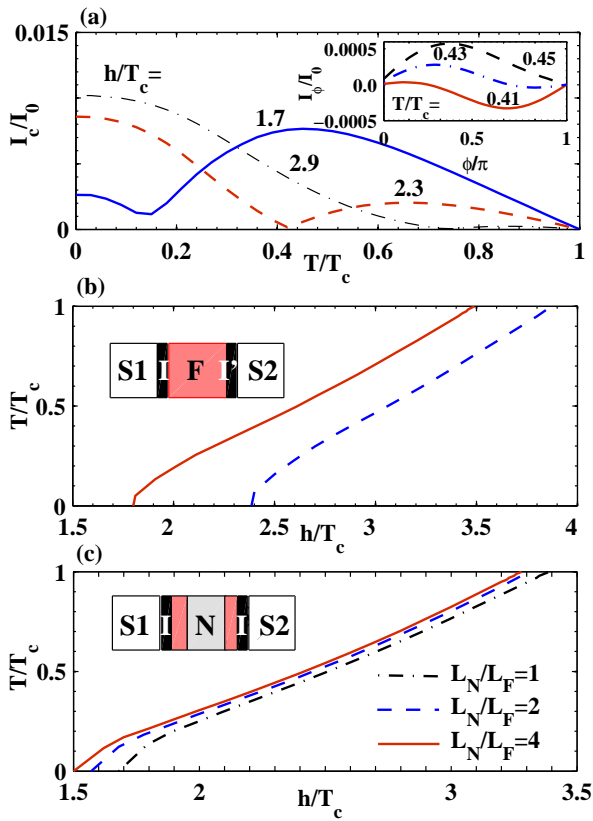}
\end{center}
\caption{ \label{Fig:3}(a) (Color online) Normalized critical
current, $I_{c}/I_{0}$, versus temperature of a symmetric SIFIS
structure with $g_{S1F}=g_{FS2}=0.018g_{T}$ for different
$h/T_{c}$, when $L_F/\xi_s=1.5$. The inset shows the corresponding
$I-\phi$ characteristic for $h/T_{c}=2.0$ in the vicinity of the
$0-\pi$ transition temperature. (b) Phase diagram of $0-\pi$
transition for a symmetric SIFIS structure with
$g_{S1F}=g_{FS2}=0.018g_{T}$ (solid line) and the corresponding
asymmetric structure (dashed line) with
$g_{S1F}=0.1g_{FS2}=0.1g_{T}$, when $L_F/\xi_s=1.5$. (c) The same
as (b) but for a symmetric SIFNFIS structure of
$g_{S1F}=g_{FS2}=0.018g_{T}$ with $L/\xi_s=1.5$ and various
$L_N/\L_F$.}
\end{figure}

Next, we examine the effect of putting insulating barriers at
FS-interfaces. In Fig.~\ref{Fig:3}a, the typical 0-$\pi$ transitions
for SIFIS structure with $L/\xi_{s}=1.5$ is shown. As one can see,
the presence of barriers adjusts the nonzero minimum cusp appearing
in the diagram of the critical current versus temperature.
Figure.~\ref{Fig:3}b manifests the $0-\pi$ phase diagram for the
symmetric SIFIS (solid curve) and the asymmetric SIFS (dashed curve)
double barrier F$_{c}$. Compared to the corresponding SFS with
$\Delta h/h_{in}=0.2$, these structures show wider transitions with
$\Delta h/h_{in}=0.93$ for SIFIS of the conductance of the
barrier $g_{S1F}=g_{FS2}=0.018g_{T}$, and $\Delta h/h_{in}=0.61$ for
SIFS of $g_{S1F}=0.1g_{FS2}=0.1g_{T}$.

We have found that the strength of the barriers between the FS
junctions is the most important parameter for determining the
width of $0-\pi$ transitions. On the one hand, as the barriers get
stronger the width of transitions becomes wider. This widening
will be more pronounced for short length structures. On the other
hand, for these structure the transition will start from a lower
exchange field in comparison with the corresponding SFS systems.

Considering the effect of the relative values of the conductances of
the two barrier, a symmetric SIFIS structures shows broader
transitions as compared to the asymmetric SIFS structure with the
same total conductance, as can be seen in Fig.~\ref{Fig:3}b.

In addition, considering the displacement of the barrier in a
S1I$_{1}$F$_{1}$I$_{2}$F$_{2}$S2 hybrid structures, we have found
that the effect of barriers becomes more important as the barriers
are closer to the ends of F$_{c}$, so that, SIFIS is the most optimal
structure regarding the width of the transitions.

Finally, we have investigated the width of the $0-\pi$ transitions for
SIFNFIS structures. The phase diagrams are shown in Fig.~\ref{Fig:3}c
for junctions with $L/\xi_{s}=1.5$ and various values of the length of
the N part, $L_{N}$. We see that, putting a normal metal between the
ferromagnets while keeping the magnetization of the system constant
increases the width of the transition somewhat. This can be
due to stronger penetration of superconductivity near the FS
boundaries where the density of magnetization is larger which
strengthens the mean effect of exchange field.

We have also observed that increasing $L_{N}$ leads to a further
increase in the width of transition. However, this increase is
saturated at higher lengthes. While the width for SIFNFIS structures
of $L_{N}=2L_{F}$ is almost doubled compared to the SIFS structure,
it is increased only few percent by increasing $L_{N}$ from $2L_{F}$
to $4L_{F}$ (see Fig.~\ref{Fig:3}c).

It is worth to note that taking the absolute width $\Delta h$
as measure of the feasibility to detect the temperature-induced
$0-\pi$ transition will lead to similar results as those of obtained
above by considering the relative width $\Delta h/h_{in}$. However,
the definition by $\Delta h/h_{in}$ seems to be more appropriate,
since higher feasibility of detection requires not only larger
$\Delta h$, but also smaller $h_{in}$ in order to have weaker
exchange-induced suppression of the supercurrent.

%
\section{\label{sec:level3}Conclusion}
In conclusion, we have investigated the width of $0-\pi$ transitions
for various diffusive ferromagnetic Josephson structures (F$_{c}$)
made of feromagnetic (F) and normal metal (N) layers and the
insulating barrier (I) contacts. We have performed numerical
calculations of the Josephson current within the quantum circuit
theory technique which takes into account fully nonlinear proximity
effect. The resulting phase diagram of $0$ and $\pi$ Josephson
couplings in the plane of $T/T_{c}$ and $h/T_{c}$ shows that the
existence of the insulating barrier contacts and the normal metal
inter-layer leads to the enhancement of the relative width of the
temperature induced transition. The relative width is parameterized
by the ratio $\Delta h/h_{in}$ with $\Delta h$ and $h_{in}$ being,
respectively, the exchange field interval upon which the transition
is possible and the initial value of $h$ at which the transition
occurs at $T=0$. We have also observed that while the conductance,
the magnetization, and the length of the F$_{c}$ junction are kept
fixed, symmetric structures with the same barrier contacts and the
same F layers in a SIFNFIS structure show larger relative width of
the transition compared to that of the asymmetric structures. Among
the studied structures, a symmetric SIFNFIS junction have the
highest $\Delta h/h_{in}$, which makes it more practicable for
highly sensitive detection of the temperature-induced $0-\pi$
transition.

\begin{acknowledgments}
M.Z. thanks W.B. for the financial support and hospitality during
his visit to University of Konstanz. W.B. acknowledges financial
support from the DFG through SFB 767 and the Landesstiftung
Baden-W\"{u}rttemberg through the Network of Competence
\textit{Functional Nanostructures}.
\end{acknowledgments}


\end{document}